\documentclass[aps,prb,superscriptaddress,twocolumn,floatfix,amsfonts,amssymb,amsmath]{revtex4-1}
\usepackage{siunitx,mhchem,bm}
\usepackage{graphicx}
\usepackage{epstopdf}

\begin{document}
	
\title{Torque magnetometry of an amorphous-alumina/strontium-titanate interface}

\author{S.L. Tomarken}
\affiliation{Department of Physics, Massachusetts Institute of Technology, Cambridge, Massachusetts 02139, USA}

\author{A.F. Young}
\affiliation{Department of Physics, Massachusetts Institute of Technology, Cambridge, Massachusetts 02139, USA}

\author{S.W. Lee}
\altaffiliation[Present address: ]{Department of Physics and Division of Energy Systems Research, Ajou University, Suwon 443-749, Korea}
\affiliation{Department of Chemistry and Chemical Biology, Harvard University, Cambridge, Massachusetts 02138, USA}

\author{R.G. Gordon}
\affiliation{Department of Chemistry and Chemical Biology, Harvard University, Cambridge, Massachusetts 02138, USA}

\author{R.C. Ashoori}
\email{ashoori@mit.edu}
\affiliation{Department of Physics, Massachusetts Institute of Technology, Cambridge, Massachusetts 02139, USA}

\begin{abstract}
We report torque magnetometry measurements of an oxide heterostructure consisting of an amorphous \ce{Al2O3} thin film grown on a crystalline \ce{SrTiO3} substrate (\mbox{\textit{a}-AO/STO}) by atomic layer deposition. We find a torque response that resembles previous studies of crystalline \ce{LaAlO3}/\ce{SrTiO3} (LAO/STO) heterointerfaces, consistent with strongly anisotropic magnetic ordering in the plane of the interface. Unlike crystalline LAO, amorphous \ce{Al2O3} is nonpolar, indicating that planar magnetism at an oxide interface is possible without the strong internal electric fields generated within the polarization catastrophe model. We discuss our results in the context of current theoretical efforts to explain magnetism in crystalline LAO/STO.
\end{abstract}

\maketitle

The interfacial two-dimensional electron gas (2DEG) in \ce{LaAlO3}/\ce{SrTiO3} (LAO/STO) heterostructures has attracted intense experimental and theoretical attention due to the hope of tailoring the properties of strongly correlated electrons via confinement\cite{Mannhart:2008,Mannhart:2010}. In particular, the polar catastrophe model that is thought to underlie the existence of a conducting two dimensional interface may provide a path to creating confined, high carrier density electron systems without the need for chemical doping\cite{Ohtomo:2004,Nakagawa:2006}. An amorphous \ce{Al2O3} thin film grown on a crystalline \ce{SrTiO3} substrate (\mbox{\textit{a}-AO/STO}) provides a useful comparison system for the role of the polar catastrophe in LAO/STO: Chemically similar yet structurally distinct, the \mbox{\textit{a}-AO/STO} interface is nonpolar. Theoretical explanations of properties common to both systems, then, should be reconciled with the absence of the strong electric fields generic to polar catastrophe models in \mbox{\textit{a}-AO/STO}. For example, similarly to LAO/STO, \mbox{\textit{a}-AO/STO} interfaces show finite electrical conductivity above a critical thickness\cite{Lee:2012}. However, the charge carrier densities are roughly one order of magnitude lower in \mbox{\textit{a}-AO/STO} than in LAO/STO, and the charge transport can be permanently suppressed by oxygen postannealing of the amorphous samples\cite{Lee:2012,Lee:2013}. These disparate observations allow a reconciliation of the apparent similarity: While it is likely that oxygen vacancy doping contributes to the charge transport in both systems, the polar electric fields in LAO/STO appear to generate a higher charge carrier density and more robust interfacial conduction.

Among the notable features of the LAO/STO interface has been the observation of magnetism \cite{Brinkman:2007,Huijben:2009,Arriando:2011,Dikin:2011,Li:2011,Bert:2011,Fitzsimmons:2011,Lee:2013b,Salluzzo:2013}. As neither of the parent materials is magnetic, the magnetism appears to be a new property generated by the electronic confinement. Relatively little is known about the origin of the magnetism, with several distinct theoretical proposals awaiting experimental testing. These proposals fall into two categories, being based either on intrinsic properties of the 2DEG in the polar catastrophe scenario\cite{Michaeli:2012,Banerjee:2013} or on (extrinsic) defect states that may or may not be related to the strong fields generated in the polar interfaces\cite{Pavlenko:2012,Fidkowski:2013,Yu:2014}. Here, we show that magnetism qualitatively similar to that observed in LAO/STO interfaces is also present in \mbox{\textit{a}-AO/STO} heterostructures. Our results raise the possibility that growth defects play a large role in the observation of magnetism in LAO/STO.

\begin{figure*}[htp]
	\centering
	\includegraphics[scale=1]{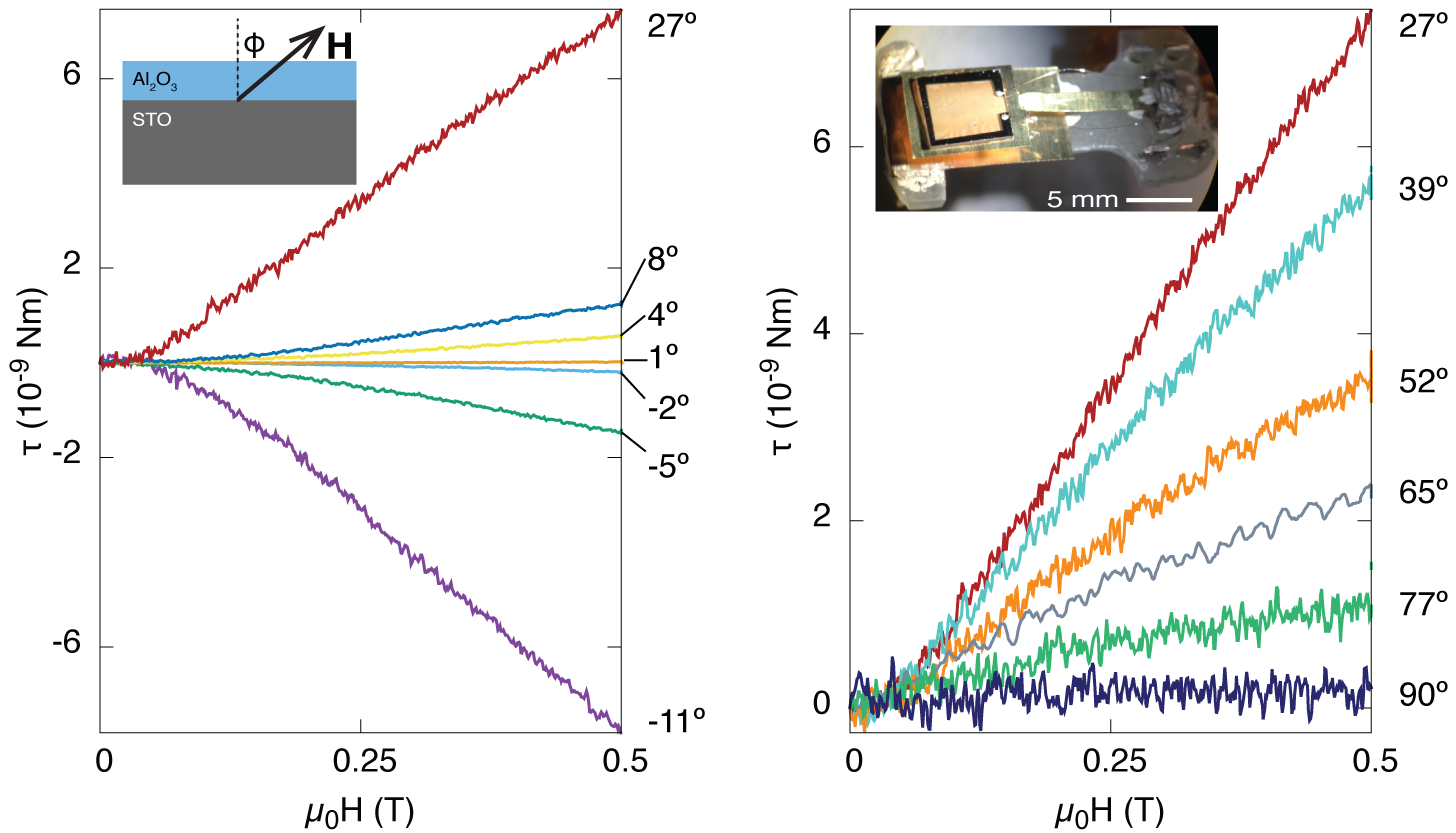}
	\caption{(Color online) Torque signal for different angles. (a) Small $\phi$.  The torque traces increase in amplitude monotonically with increasing angle $\left|\phi\right|$. The torque traces invert about $\phi=0^\circ$. Inset: Schematic depicting the interface and applied field. (b) Large $\phi$.  For $\phi > 27^\circ$, the torque signal monotonically decreases with increasing $\phi$. The torque traces are proportional to $H_\perp = H\cos\phi$. As $\phi$ approaches $90^\circ$, the torque amplitude vanishes. Inset: Picture of a brass cantilever with the sample. The dark ring around the sample is a calibration loop.}
	\label{torque}
\end{figure*}

Our measurement involves placing a heterostructure on the end of a brass cantilever so that the interface is parallel to the cantilever [Fig.~\ref{torque}(b) inset]. In the presence of an applied magnetic field $\bm{H}$ the cantilever will experience a torque $\bm{\tau} = \bm{m} \times \mu_0\bm{H}$. For a moment $\bm{m}$ which is in the plane of the interface, the torque signal will be $\tau = \mu_0 m(H)H\cos\phi$, where $\phi$ is the angle between the applied magnetic field and the axis perpendicular to the interface [Fig.~\ref{torque}(a) inset]. Such a torque will deflect the cantilever, and this deflection can be detected via the change in the cantilever's capacitance to a nearby conducting plane. We point out that torque magnetometry is directly sensitive to the magnetic moment parallel to the plane of the cantilever. Accordingly, we are sensitive to planar magnetic contributions from both the interface as well as the substrate. However, we find it unlikely that magnetic contributions far from the interface would show strong planar anisotropy. We note that torque measurements performed on bare STO substrates in Ref.~\onlinecite{Li:2011} showed no signs of planar magnetism.

Our sample consists of amorphous alumina grown on a single crystal STO substrate using an atomic layer deposition technique described previously \cite{Lee:2012}. $\SI{5}{\nano\meter}$ of amorphous alumina were grown on a \ce{TiO2}-terminated STO substrate at $\SI{300}{\celsius}$ using trimethylaluminum and \ce{H2O} as the aluminum precursor and oxygen source, respectively. The room temperature electron density at the \textit{a}-AO/STO interface was $\SI{3e12}{\per\centi\meter\squared}$ (determined by Hall measurement). A $\SI{100}{\nano\meter}$ thick aluminum loop was subsequently deposited around the edge of the $\SI{6}{\milli\meter}\times\SI{6}{\milli\meter}$ sample [see Fig.~\ref{torque}(b) inset] to allow \textit{in situ} calibration of the cantilever torque constant. The loop was grounded when measuring the heterostructure torque. The sample was fixed with GE varnish to a cantilever made from $\SI{25}{\micro\meter}$ thick brass foil. The cantilever was suspended on a glass post above a fixed brass conducting plane on a G10 chip carrier.

We measured capacitance with a $\SI{5}{\volt}$ excitation at $\SI{8}{\kilo\hertz}$ using a General Radio $1615$A capacitance bridge and lock-in amplifier. Measurements were performed with the sample immersed in liquid \ce{^3He} (for base temperature measurements) or \ce{^3He} gas (for the high temperature measurements in Fig.~\ref{temp}).  The magnetic field was ramped at a constant rate of $\SI{0.3}{\tesla\per\minute}$ with no detectable hysteresis. We report data from one sample with our most complete torque series and temperature dependence. We also saw similar signs of magnetic ordering in one other nominally identical sample.    

Figure \ref{torque} shows torque traces for different interface orientations at $\SI{400}{\milli\kelvin}$. The angle $\phi$ describes the tilt angle of the magnetic field with respect to the axis perpendicular to the interface [Fig.~\ref{torque}(a) inset]. For a constant in-plane moment $m(H) = m_0$, the torque signal is linear in $H$. For a more general in-plane moment $m(H)$ which evolves as a function of the applied field, the torque traces will be nonlinear at low applied field but will reach a linear regime once the in-plane moment has saturated (as in a superparamagnetic or ferromagnetic system). In addition to a possible linear signal indicative of magnetic ordering, the sample may also have an overall paramagnetic or diamagnetic background. Although torque of the form $\textbf{m}\times \mu_0\textbf{H}$ vanishes for paramagnetic and diamagnetic contributions ($\textbf{m} \propto \textbf{H}$), the presence of small magnetic field gradients perpendicular to the plane of the cantilever results in a deflection proportional to $\bm{\nabla}\left(\textbf{m}\cdot\mu_0\textbf{H}\right)$. For paramagnetic or diamagnetic moments ($\textbf{m} \propto \textbf{H}$) this corresponds to a deflection proportional to $H^2$. At low applied field we are dominated by the linear deflection response, which we interpret (following Ref.~\onlinecite{Li:2011}) as arising from in-plane magnetic ordering. We restrict our range to $|\mu_0H| < \SI{0.5}{\tesla}$ to remain within this regime. (See Fig.~\ref{temp} for an example of a larger field range with quadratic contributions.)

\begin{figure}[h!]
	\centering
	\includegraphics[scale=1]{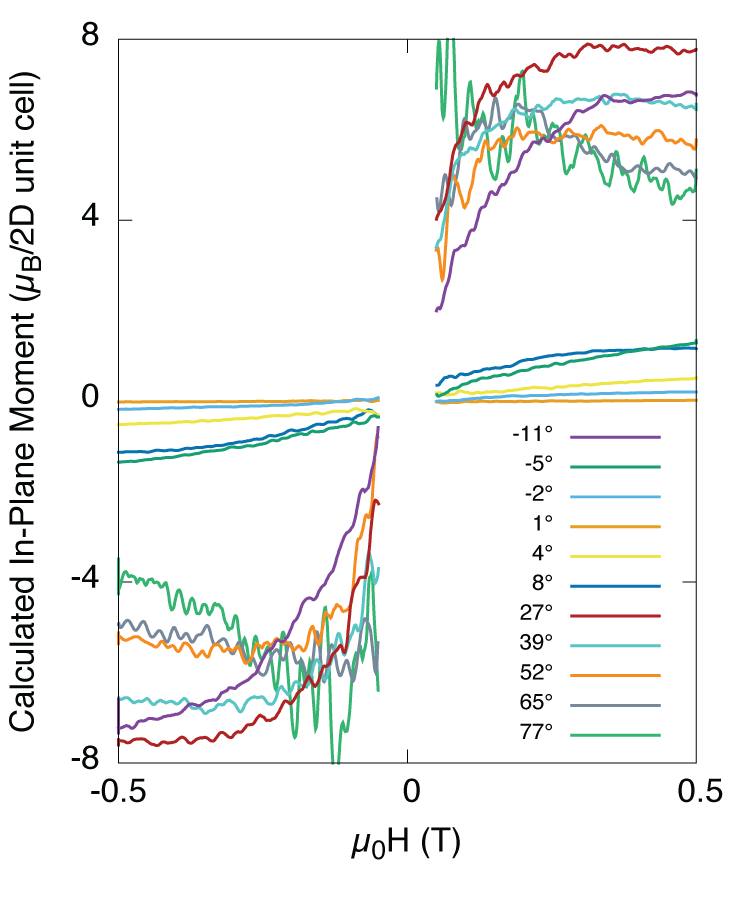}
	\caption{(Color online) Magnetic moment calculation. The magnetic moment $m(H)$ was calculated from the torque traces by dividing by the perpendicular applied field: $m(H) = \tau(H)/\mu_0H\cos\theta$. The trace corresponding to $\phi = 90^\circ$ was excluded due to the loss of sensitivity around $\phi \sim 90^\circ$. Data below $\SI{50}{\milli\tesla}$ have been suppressed due to loss of torque sensitivity at low field. The net magnetic moment has been expressed in units of Bohr magnetons $\mu_\text{B}$ per 2D STO unit cell (u.c.).}
	\label{moment}
\end{figure}

\begin{figure}[b!]
	\centering
	\includegraphics[scale=1]{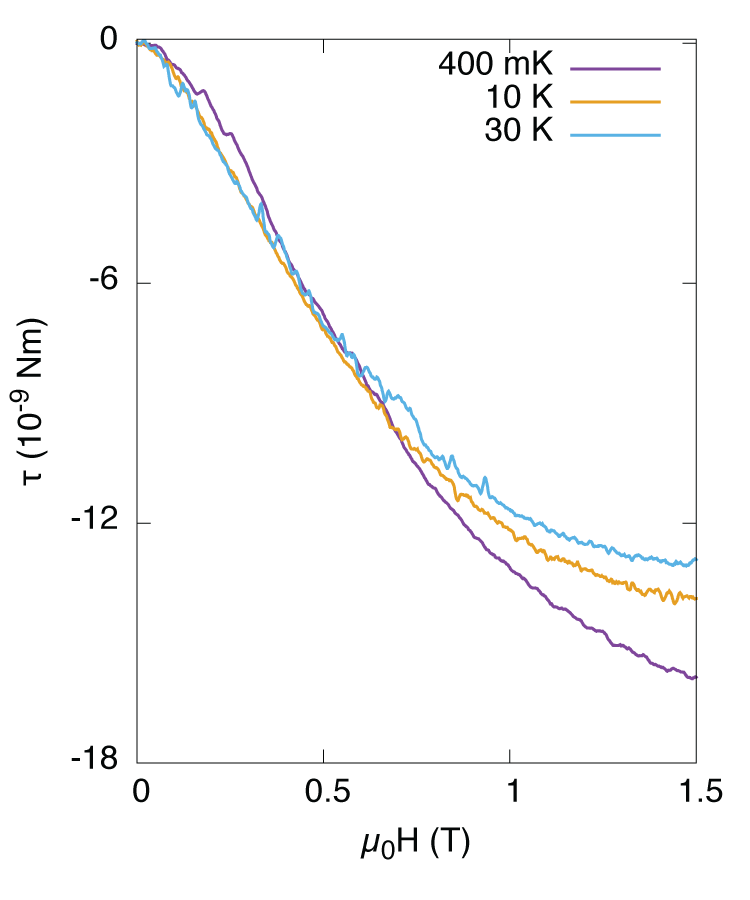}
	\caption{(Color online) Temperature dependence of torque. Torque traces at constant $\phi = -11^\circ$ from $\SI{400}{\milli\kelvin}$ to $\SI{30}{\kelvin}$. There is no quantitative change in torque signal below $\SI{0.5}{\tesla}$ throughout the accessible temperature range. The curvature at the higher field is due to an overall diamagnetic background.}
	\label{temp}
\end{figure}

Figure \ref{torque}(a) shows torque traces for small $\phi$. At $\phi \approx 0^\circ$ there is a negligible torque signal, indicating the absence of an in-plane magnetic moment. As the angle is tilted in either the positive or negative direction, the torque signals monotonically increase in magnitude with the sign of the slope determined by the tilt direction. The slope inversion about $\phi=0^\circ$ suggests that, as in LAO/STO\cite{Li:2011}, there is either no coercive field or one that cannot be resolved with our technique, which loses sensitivity at small applied field. The increase in torque signal indicates an increase of the in-plane moment up to a maximum value which occurs between about $|\phi| \approx 10^\circ$ and $25^\circ$. At applied field $H \gtrsim \SI{0.25}{\tesla}$, the torque traces are linear, indicating a constant in-plane moment consistent with either superparamagnetism or ferromagnetism with a small coercive field which we cannot detect. We note that the size of the in-plane moment in the linear regime shows dependence on the out-of-plane field (see Fig.~\ref{moment}). This is consistent with a picture of the interfacial moments canting slightly out of the plane of the interface when subjected to a strong out-of-plane field. However, because our measurement is only sensitive to the in-plane field, we have no way of verifying this picture and make no claims as to the origin of this dependence. Figure \ref{torque}(b) shows torque traces at larger tilt angles. After passing through a maximum at $\phi = 27^\circ$, the torque signal monotonically decreases as $\phi$ approaches $90^\circ$, where the applied field is parallel to the interface. Because $\tau \propto H_\perp$, the torque signal is strongly suppressed at large angles and vanishes completely at $\phi = 90^\circ$.

The angle between the applied field and the interface was determined by first setting the interface perpendicular to the applied magnetic field ($\phi = 0^\circ$). This was achieved by both observing a torque trace with zero amplitude as well as a maximum in the cantilever to conducting plane capacitance. Our cryogenic rotation stage was controlled by a room temperature calibrated linear actuator. By referencing the position of the linear actuator relative to $\phi = 0^\circ$ we could determine the angle of inclination of the rotation stage with respect to the applied field. However, this does not take into account the mechanical bending of the cantilever as the rotation stage changes orientation. This may cause an error in our angle calibration by as much as $5^\circ-10^\circ$, particularly at large angles. However, we stress that this calibration does not affect the essential results of our measurement.  

Comparison of the torque data in Fig.~\ref{torque} with Fig.~4 from Ref.~\onlinecite{Li:2011} shows that our data is in qualitative agreement with previous results from crystalline LAO/STO. However, the torque signal from \mbox{\textit{a}-AO/STO} is roughly one order of magnitude stronger than that observed in Ref.~\onlinecite{Li:2011} despite a similar sample size, indicating a much higher net in-plane moment in the amorphous system.  We can calculate the size of the in-plane magnetic moment through the relation $m(H) = \tau(H)/(H\cos\phi)$. Figure \ref{moment} shows $m(H)$ traces calculated from selected torque traces in Fig.~\ref{torque}. The magnetic moment has been expressed in units of the Bohr magneton and normalized by the number of 2D unit cells (u.c.) in one layer of crystalline STO. Note that this normalization corresponds to the effective areal moment density, as if all of the signal were confined to the interface. Although we do not know the true spatial distribution of the magnetic moments, this convention allows comparison of our experimental data to theories making quantitative predictions of local moments within the first interfacial \ce{TiO2} layer as well as a direct comparison to Ref.~\onlinecite{Li:2011}. Explicit computation of $\tau(H)/(\mu_0H\cos\phi)$  within the $|\mu_0H|< \SI{50}{\milli\tesla}$ range has been suppressed due to the loss of torque sensitivity at low field. We have also excluded the trace corresponding to $\phi=90^\circ$ due to loss of torque sensitivity when the applied field is parallel to the interface. The in-plane moment achieves a maximum of $5-8\,\mu_\text{B}/\text{2D u.c.}$ in this particular sample and is roughly one order of magnitude larger than the $0.3-0.4\,\mu_\text{B}/\text{2D u.c.}$ estimate of crystalline LAO/STO in \textcite{Li:2011} [see Fig.~1(d) in Ref.~\onlinecite{Li:2011}]. However, the large moment density makes it unlikely that the magnetic signal arises from a truly two-dimensional layer at the interface. The magnetic moments are likely distributed in a three-dimensional volume within the STO. Lacking another obvious mechanism to explain the magnetic anisotropy, we suspect the magnetic moments are near the interface. 

We also measured the temperature dependence of the magnetism. In Fig.~\ref{temp} we plot the torque $\tau(H)$ traces at constant tilt angle $\phi = -11^\circ$ from $\SI{400}{\milli\kelvin}$ up to $\SI{30}{\kelvin}$, the highest temperature achievable in the \ce{^3He} cryostat. There is no quantitative change in the torque signals below $\SI{0.5}{\tesla}$ (the regime most sensitive to magnetic ordering). The curvature in the torque traces at a higher applied field is due to the overall diamagnetic background of the sample and shows a weak temperature dependence. \textcite{Li:2011} found a similar lack of temperature dependence up to $\SI{40}{\kelvin}$ (see Fig.~3 in Ref.~\onlinecite{Li:2011}).

Our amorphous sample demonstrates magnetic ordering with strong planar anisotropy, no detectable coercive field, and stability at elevated temperatures --- in striking similarity with results from crystalline LAO/STO. However, our estimate of $5-8\,\mu_\text{B}/\text{2D u.c.}$ differs quantitatively from the magnetic moment estimate of $0.3-0.4\,\mu_\text{B}/\text{2D u.c.}$ found in Ref.~\onlinecite{Li:2011}. Because the \mbox{\textit{a}-AO/STO} interfaces should have much a smaller electric polarization than crystalline LAO/STO, yet show a much larger areal moment density, our data raise the possibility that growth defects are largely responsible for magnetic ordering in both epitaxial and amorphous oxide interfaces which have not been oxygen postannealed. 

In addition to considering previous torque measurements, our results may provide an important context for recent reports of interfacial magnetism in LAO/STO. We note that Ref.~\onlinecite{Salluzzo:2013} found evidence of the absence of interfacial magnetism in oxygen postannealed LAO/STO heterostructures using x-ray absorption spectroscopy. The residual magnetism observed in oxygen annealed samples \cite{Salluzzo:2013} was consistent with previous neutron reflectometry results \cite{Fitzsimmons:2011}. However, Ref.~\onlinecite{Lee:2013b} found evidence of \ce{Ti^{3+}} magnetism in oxygen annealed samples. It has also been recently proposed that the polarization catastrophe may generate interfacial defects which ultimately are responsible for localized magnetic moments at the interface\cite{Yu:2014}. In this scenario \ce{Ti}-on-\ce{Al} antisite defects are responsible for magnetism and not oxygen vacancies, as proposed elsewhere\cite{Pavlenko:2012}. Our data do not rule out the possibility of distinct mechanisms driving qualitatively similar magnetism in epitaxial LAO/STO and \mbox{\textit{a}-AO/STO}. Rather, our results point to the necessity of incorporating the role of growth related defects when describing interfacial magnetism in oxides.

There are several future measurements that could isolate which aspects of the oxide interface landscape are responsible for magnetism. Namely, performing similar torque measurements on oxygen annealed samples could demonstrate whether or not the magnetism arises from oxygen vacancies. Similarly, performing magnetometry on \textit{a}-AO/STO structures with different overlayer thicknesses (especially near the \SI{1.5}{\nano\meter} critical thickness) could reveal to what degree (if any) the magnetism is influenced by the presence of a 2DEG. However, as in the crystalline system, isolation of the role of oxygen vacancies from electronic reconstructions remains a challenge.  

This work was sponsored by the BES Program of the Office of Science of the U.S. DOE, Contract No.~FG02-08ER46514, and the Gordon and Betty Moore Foundation, through Grant No.~GBMF2931. This material is based upon work supported by the National Science Foundation Graduate Research Fellowship under Grant No.~1122374. A portion of this work was performed at the National High Magnetic Field Laboratory, which is supported by National Science Foundation Cooperative Agreement No.~DMR-1157490, the State of Florida, and the U.S. Department of Energy. A.F.Y.~acknowledges the support of the MIT Pappalardo Fellowship in Physics.

\bibliography{amoroxideArXiv}
\end{document}